\documentclass[conference]{IEEEtran}
\IEEEoverridecommandlockouts
\usepackage{fancyhdr}
\usepackage{amssymb}
\usepackage{amsmath}

\usepackage{amsfonts}
\usepackage{algorithmic}
\usepackage{algorithm}
\usepackage{array}
\usepackage[caption=false,font=normalsize,labelfont=sf,textfont=sf]{subfig}
\usepackage{textcomp}
\usepackage{stfloats}
\usepackage{url}
\usepackage{verbatim}
\usepackage{graphicx}
\usepackage{mathtools}
\usepackage{siunitx}
\usepackage[hidelinks]{hyperref}
\usepackage{wrapfig}
\usepackage{tabularx}
\usepackage{amsthm}
\usepackage{booktabs}

\def\BibTeX{{\rm B\kern-.05em{\sc i\kern-.025em b}\kern-.08em
    T\kern-.1667em\lower.7ex\hbox{E}\kern-.125emX}}

\def\BibTeX{{\rm B\kern-.05em{\sc i\kern-.025em b}\kern-.08em
    T\kern-.1667em\lower.7ex\hbox{E}\kern-.125emX}}

\usepackage{titlesec}

\usepackage{amsmath}

\renewcommand{\thesection}{}  
\renewcommand{\thesubsection}{} 
\renewcommand{\thesubsubsection}{}

\titleformat{\section}[block]{\bfseries\normalsize}{\thesection}{0pt}{}  
\titleformat{\subsection}[block]{\itshape\normalsize}{\thesubsection}{0pt}{}  
\titleformat{\subsubsection}[block]{\itshape\normalsize}{\thesubsubsection}{0pt}{} 

\begin{document}
\theoremstyle{definition}
\newtheorem{definition}{Definition}

\theoremstyle{assumption}
\newtheorem{assumption}{Assumption}
\rmfamily

\title{{\fontsize{16}{19}\selectfont \textbf{Safe Trajectory Sets for Online Operation of Power Systems under Uncertainty}}}

\author{
    \IEEEauthorblockN{
        Florian Klein-Helmkamp\IEEEauthorrefmark{1},
        Tina Möllemann\IEEEauthorrefmark{1},
        Irina Zettl\IEEEauthorrefmark{1},
        Steffen Kortmann\IEEEauthorrefmark{1}\IEEEauthorrefmark{2},
        and Andreas Ulbig\IEEEauthorrefmark{1}\IEEEauthorrefmark{2}\\}
    \IEEEauthorblockA{
        \IEEEauthorrefmark{1}IAEW at RWTH Aachen University, Aachen, Germany \\
        \IEEEauthorrefmark{2}Fraunhofer Center Digital Energy, Fraunhofer FIT, Aachen ,Germany
        \\\{f.klein-helmkamp,i.zettl,s.kortmann,a.ulbig\}@iaew.rwth-aachen.de
        \\ tina.moellemann@rwth-aachen.de
    }}
\maketitle
\thispagestyle{fancy}

\begin{abstract}
Flexibility provision from active distribution grids requires efficient and robust methods of optimization and control suitable to online operation. In this paper we introduce conditions for the secure operation of power systems using feedback optimization based controllers. The feasibility of system operation is defined using the Feasible Operating Region (FOR), which serves as a boundary for safe system states. We propose a method to compute safe trajectory sets by projecting the high-dimensional system state onto the two-dimensional PQ-plane, enabling intuitive analysis of controller behavior. Our method is demonstrated on an exemplary sub-transmission system, where we evaluate the controller's performance and robustness under disturbances and uncertainties. Results highlight the suitability of the proposed approach for assessing reachable and feasible system states in cases where the controlled grid is subjected to disturbances.
\end{abstract}


\section{Introduction}
Distribution grids are increasingly taking on an active role in power system operations, including the provision of ancillary services by distributed energy resources (DER), aggregators, or distribution system operators (DSOs). This includes frequency control \cite{Karagiannopoulos_2020,Stanojev_2021}, voltage support for the overlaying grid \cite{Karagiannopoulos_2021,Escobar_2022,Ortmann_2023a}, and congestion management on transmission system level \cite{Alizadeh_2022,Contreras_2021}. The involvement of potentially large numbers of DERs and subsystems necessitates efficient and robust control and optimization methods tailored to the specific ancillary services. Given the high security requirements in power systems, a potential control architecture must be validated for performance and robustness, even under uncertainty, disturbances, and noise. To address these challenges of online control, there has been growing interest in the application of closed-loop techniques for solving optimization problems in power systems \cite{Molzahn_2017,Häberle_2020}. These methods leverage the advantages of feedback control by incorporating real-time measurements while solving an optimization problem, leading to an iterative trajectory of systems states tracking the optimal solution. This approach offers several advantages, including robustness to model mismatch, resilience to disturbances, and lower computational complexity compared to conventional optimization problems such as the AC optimal power flow.
\begin{figure}[tb]
    \centering
    \includegraphics[width=1\linewidth]{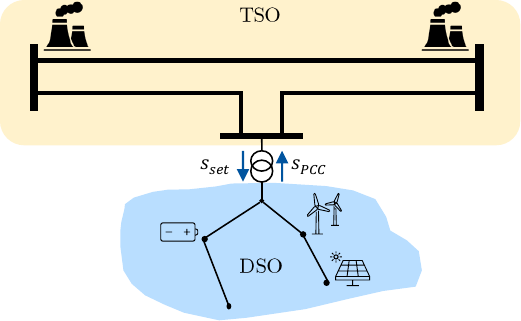}
    \caption{Hierarchical flexibility provision at the TSO-DSO interface.}
    \label{fig:tso_dso}
\end{figure}

\subsection{Related Work}
\label{subsec:Related Work}
Incorporating distribution grid flexibility in online grid operation requires controlling the load flow across the point of common coupling (PCC) to the superimposed grid layer by dispatching new operating points to individual DER. Several approaches for the vertical coordination of flexibility in grid operation have been proposed so far. Going beyond the dispatch, methods for the quantification of safe system states for distribution grids in context of flexibility provision were described in recent years. To further distinguish the work presented in this paper and to put it in context regarding the current state of research, we introduce four relevant categories for our literature review. Those include the \emph{Quantification} of flexibility and safe operational regions for distribution grids, \emph{Robustness} of dispatch and control of flexibility, an emphasis on \textit{Online Operation} and finally the application of \emph{Closed-Loop} techniques for control of flexibility. An overview of all analysed references is shown in \autoref{tab:lit_review}.

\begin{table}[tb]
    \centering
    \caption{Classification of examined literature by research focus.}
    \label{tab:lit_review}
    \begin{tabularx}{\columnwidth}{XXXX}
        \toprule
         \textbf{Flex. Quant.} & \textbf{Robustness} & \textbf{Online Op.} & \textbf{Closed Loop} \\
        \midrule
         \cite{Karagiannopoulos_2020,Alizadeh_2022,Contreras_2021,Papazoglu_2022,Givisiez_2020,Uzum_2024,Silva_2018,Lopez_2021,Patig_2022,Bandeira_2024,Früh_2022,Kolster_2022,Sarstedt_2022}& \cite{Patig_2022,Früh_2022,Kolster_2022,Hauswirth_2024}  & \cite{Stanojev_2021,Karagiannopoulos_2021,Escobar_2022,Ortmann_2023a,Molzahn_2017,Früh_2022,Picallo_2020,Ortmann_2020,Picallo_2023,Bernstein_2019,Zhan_2024,Ortmann_2023,Ortmann_2024,KH_2023,KH_2024,Zettl_2024} & \cite{Stanojev_2021,Ortmann_2023a,Molzahn_2017,Häberle_2020,Picallo_2020,Bolognani_2015,Ortmann_2020,Picallo_2023,Bernstein_2019,Zhan_2024,Hauswirth_2024,Ortmann_2023,KH_2024,Ortmann_2024,Zettl_2024}\\
        \bottomrule
    \end{tabularx}
\end{table}

Distribution grid flexibility is mainly discussed taking one of two perspectives. First, the quantification of available flexibility is of interest to allow for a consideration in operational planning \cite{Papazoglu_2022}. Second, the dispatch of a flexibility request requiring  coordination of multiple, often numerous DER \cite{Givisiez_2020,Uzum_2024}.

\paragraph{Quantification of Flexibility}
A common method to quantify the available flexibility is the feasible operating region (FOR), i.e. the set of all complex load flows at the coupling transformer between two grid layers that are feasible subject to the constraints of the system \cite{Silva_2018}. An approach to quantify aggregated flexibility on distribution grid level for the use case of congestion management in superimposed grid layers is described in \cite{Contreras_2021}. The authors propose an approach to determine the FOR based on a linear OPF. Another method to determine the FOR based on relaxation of the power flow equations is described in \cite{Lopez_2021}. The authors propose an algorithm constructing the FOR and compare the results for different formulations of the OPF. Extending on optimization based approaches to determine the FOR, the authors of \cite{Patig_2022} introduce chance constraints to a linear approximation of the OPF to account for possible uncertainties while determining the FOR. Further tackling the feasibility of flexibility aggregation an approach based on Approximate Dynamic Programming is introduced in \cite{Bandeira_2024}. The authors use the DistFlow model to capture the physical constraints of the system, ensuring feasibility in aggregation.

\paragraph{Coordination of Flexibility}
Using flexibility in grid operation requires coordination of DER based on a specified performance metric. This is usually referred to as \emph{disaggregation}. An approach to coordinate aggregated flexibility in a cascaded manner using a linear OPF is described in \cite{Früh_2022}. Utilizing flexibility at the TSO-DSO-interface in a cost-optimal way, the authors of \cite{Sarstedt_2022} propose a disaggregation method based on MILPs with respect to the individual cost of flexibility. The authors of \cite{Kolster_2022} propose the use of robust optimization to determine and coordinate available flexibility for the use case of curative system operation under consideration of uncertainty. All of the approaches to flexibility coordination mentioned so far are based on solving optimization problems in grid operation offline. This is a well established practice in the domain of power systems, but it lacks robustness and can be computationally expensive when AC load flow equations are included in the optimization problem.

\paragraph{Online Feedback Optimization}
Power system optimization incorporating real-time feedback has been proposed for a variety of different use-cases. An approach to track the solution of the AC optimal power flow on the physical power system is presented in \cite{Picallo_2020}. Approaches usually entail solving optimization problems by iteratively steering the physical system to an optimal solution based on linear input output sensitivities \cite{Bolognani_2015,Hauswirth_2024}. These can either be determined offline \cite{Ortmann_2020} or learned from measurement \cite{Picallo_2023}. An approach to real-time coordination of DER on distribution grid level according to specified performance objectives, such as set point tracking at the substation is described in \cite{Bernstein_2019}. The authors of \cite{Zhan_2024} describe an approach to voltage control on DSO level for different time scales. From a systemic perspective the application of Online Feedback Optimization (OFO) in provision of ancillary services requires the optimization and activation of such measures in online grid operation. An approach to curative grid operation using Online Feedback Optimization is introduced in \cite{Ortmann_2023}. OFO is performing supervisory control, recognizing critical system states and activating remedial actions on the same system layer if needed. To further leverage flexibility that is connected to underlying grid layers, i.e. sub-transmission and distribution systems, a hierarchical control structure for an efficient dispatch based on OFO for curative flexibility is introduced in \cite{KH_2023, KH_2024}. Different approaches to tuning the proposed control stack are introduced in \cite{Ortmann_2024} and \cite{Zettl_2024}. These works illustrate, that the performance of a control stack based on OFO depends on the chosen formulation of the optimization problem and other parameters, e.g. the gain of the controller. The application in safety critical infrastructure like power systems therefore leads to a trade-off between robustness and performance of a proposed control stack. So far previous publications on flexibility quantification, flexibility coordination and feedback-based optimization lack methods to verify a specific control stack with respect to the problem of robustness and performance.

\subsection{Main Contribution}
\label{subsec:Main Contribution}
In order address this gap the main contribution of this work is twofold. First, we introduce an extended hierarchical control architecture for multi-layer flexibility provision in online grid operation. To this end, we incorporate underlying grid layers as flexible resources to support load flow control. Second, we propose a novel method for calculating trajectory sets under uncertainty. This approach verifies the proposed control stack in terms of safety and robustness while quantifying the available flexibility based on the specific implementation and controller tuning. To achieve this, we systematically analyze trajectories resulting from adapting to new load flow set points at the point of common coupling between two grid layers. Based on the theoretical guarantees and previous studies we expect OFO to ensure robust and safe operation of power systems under uncertainty and disturbances by leveraging real-time measurements to iteratively adapt control actions, even in scenarios with limited controllability and incomplete system models.

\section{Closed-Loop Control Architecture}
\label{sec:CLC}
This section describes the proposed closed-loop control architecture in detail. The overall control stack follows the hierarchical structure of interacting grid layers, starting from transmission system to sub-transmission system to one or several distribution grids participating. We first introduce the general optimization problem describing a dispatch of flexibility within the system constraints. Afterwards, we detail the proposed solution approach to the formulated problem based on an implementation of projected gradient descent in closed-loop with the physical system.

\subsection{Flexibility Provision}

We denote $N$ as the set of buses, $B$ as the set of branches (i.e., lines and transformers), and $F$ as the set of all flexible actors in a grid. Let $\mathbf{v} \in \mathbb{R}^{|N|}$ represent the vector of bus voltages, $\mathbf{s} \in \mathbb{R}^{|B|}$ the vector of active and reactive branch flows, and $\mathbf{p} \in \mathbb{R}^{|F|}$ and $\mathbf{q} \in \mathbb{R}^{|F|}$ the vectors of active and reactive power injections for flexible actors. For a given set point of apparent power flow at the PCC between two grid layers, $\mathbf{s}_{\text{set}} = [p_{\text{set}}, q_{\text{set}}]^T$, we formulate the following dispatch problem for flexibility:

\begin{equation}
	\label{eq:dispatch problem}
	\begin{aligned}
		\min_{\mathbf{p}, \mathbf{q}} &\quad \Phi =  ||p_{\text{set}} - p_{\text{PCC}}||^{2} + ||q_{\text{set}} - q_{\text{PCC}}||^{2}  \\
		\text{s.t.}
		&\quad \mathbf{v}_{\text{min}} \leq \mathbf{v} \leq \mathbf{v}_{\text{max}}, \\
		&\quad \mathbf{s}_{\text{min}} \leq \mathbf{s} \leq \mathbf{s}_{\text{max}}, \\
		&\quad \mathbf{p}_{\text{min}} \leq \mathbf{p} \leq \mathbf{p}_{\text{max}}, \\
		&\quad \mathbf{q}_{\text{min}} \leq \mathbf{q} \leq \mathbf{q}_{\text{max}}
	\end{aligned}
\end{equation}

This optimization minimizes the Euclidean norm of the difference between the measured apparent power flow, $\mathbf{s}_{\text{PCC}} = [p_{\text{PCC}}, q_{\text{PCC}}]^T$, and the set point, $\mathbf{s}_{\text{set}} = [p_{\text{set}}, q_{\text{set}}]^T$, to achieve the desired operating point while satisfying system technical constraints. These constraints, detailed in \eqref{eq:dispatch problem}, include limits on bus voltages, $v_n \ \forall n \in N$, and branch flows, $s_i \ \forall i \in B$. Power limits for available flexibility, such as controllable loads, generation, and other grid layers, are also incorporated as inequality constraints. Since this optimization problem is solved online on the physical system, the AC power flow equations are not explicitly included as equality constraints. Instead, constraint satisfaction is ensured through real-time voltage and branch flow measurements. In the feedback optimization-based approach, the dispatch problem is solved iteratively. At each step, the cost function and constraints are incorporated into a quadratic optimization problem to determine the controller's next action. The detailed formulation of the proposed OFO controller, including its iterative implementation, is presented in the following section.

\subsection{Feedback Optimization Based Control}
\label{subsec:OFO}
\begin{figure}[tb]
	\centering
	\includegraphics{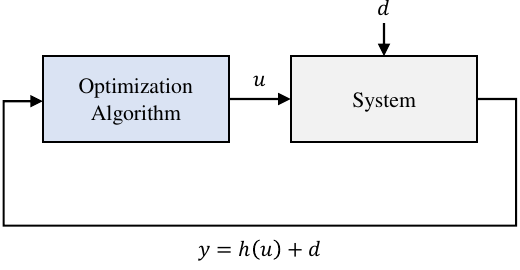}
	\caption{Closed-loop optimization on physical system considering online measurements.}
	\label{fig:basic_loop}
\end{figure}

In this section, we introduce the general concept of the OFO controller presented in this work. As stated in the introduction, dispatch problems in the form of \eqref{eq:dispatch problem} are typically solved offline for a model of the system with an estimate of possible disturbances. In contrast, the feedback-based approach iteratively steers the system by directly acquiring the relevant states through online measurements (see \autoref{fig:basic_loop}). In this work we choose projected gradient descent as the underlying optimization algorithm for the controller.

We consider a system $\mathbf{y} \in \mathbb{R}^p$ described by its inputs $\mathbf{u} \in \mathbb{R}^q$, a function mapping inputs to outputs $h(\mathbf{u})$ and disturbances $\mathbf{d} \in \mathbb{R}^p$, where:
\begin{equation}
\label{eq:system}
    \mathbf{y} = h(\mathbf{u}) + \mathbf{d}
\end{equation}
For our presented use case, this entails a flexibility-providing power system that includes both controllable and non-controllable units, e.g. loads and DERs. The dispatch of flexibility is optimized based on \eqref{eq:dispatch problem} resulting in new operating points for flexible units. After actuation, all units adjust their feed-in or consumption of active and reactive power accordingly.

To ensure stability and convergence to the optimum of the cost function and to allow for an analysis of steady states, we separate the controller's actions from the fast dynamics of the system \cite{Hauswirth_2021}. This can be achieved by selecting an appropriate cycle time between the iterations awaiting all relevant faster dynamics of the system such as frequency and voltage transients. In this way we can abstract the differential-algebraic equations of the system describing its physical behavior to their resulting steady states. We express this in the following assumption.

\begin{assumption}(Time-scale separation)
For the dynamic system $y$ time-scale separation between the actions $u$ of the central OFO controller and the fast dynamics of the system, as well as the adjustment to new set points for actuators within the system is assumed.
\end{assumption}

To drive the system to the optimal solution of the underlying optimization problem \eqref{eq:dispatch problem} we approximate the behavior of the system by its steady state map. The proposed controller utilizes an input-output sensitivity map $\nabla h$ that represents the gradient of system output $y$ with respect to control input $u$. The sensitivities are determined by linearly approximating around the smooth power flow manifold for an initial operating point. They include the relevant states, such as bus voltages, branch flows and especially power flow at the PCC. We calculate $\nabla h$ with:
\begin{equation}  
\nabla h(u) = \begin{bmatrix}
\nabla_p v(p,q) \\
\nabla_q v(p,q) \\
\nabla_p s(p,q) \\
\nabla_q s(p,q) \\
\nabla_p p_{PCC}(p,q) \\
\nabla_q p_{PCC}(p,q) \\
\nabla_p q_{PCC}(p,q) \\
\nabla_q q_{PCC}(p,q)
\end{bmatrix}
\end{equation}
Due to the inclusion of feedback in each step of the controller, $\nabla h$ is computable offline for a fixed initial operating point. We are subsequently able to use this constant approximation in online operation. For the controller to act on the system we define the vector of measurements $y \in \mathbb{R}^{p}$ with
\begin{equation}
	\label{eq:system}
	\mathbf{y} = [v_{1}, ..., v_n, s_1, ..., s_m]
\end{equation}
and the vector of set points $u\in\mathbb{R}^q$ with
\begin{equation}
	\label{eq:system}
	\mathbf{u}= [p_{1}, ..., p_j, q_1, ..., q_j].
\end{equation}
OFO acquires the system state and actuates controllable flexibility in an iterative manner. To converge to the optimal solution of the underlying optimization problem and its convex cost function $\Phi$ we use a formulation based on projected gradient descent. For a positive scalar $\alpha \in \mathbb{R}_{>0}$ we therefore determine the step-size of the next control action $\hat{\sigma}(u,y)$ by solving the quadratic problem:
\begin{equation}
\label{eq:ofo_qp}
	\begin{aligned}
		\hat{\sigma}(u,y) \coloneqq \arg \min_{w \in \mathbb{R}^{p}} \quad & \| w + 2H(u)^{T}\nabla \Phi(u,y)\|^{2}\\
		\textrm{s. t.} \quad & \begin{bmatrix}p_{\text{min,j}} \\q_{\text{min,j}}\end{bmatrix}\leq \begin{bmatrix}p_j \\q_j\end{bmatrix} + \alpha w \leq \begin{bmatrix}p_{\text{max,j}} \\q_{\text{max,j}}\end{bmatrix}\\
		\quad & v_{\text{min}} \leq v_{\text{meas}} + \alpha \nabla h(u) w \leq v_{\text{max}} \\
  		\quad & s_{\text{min}} \leq s_{\text{meas}} + \alpha \nabla h(u) w \leq s_{\text{max}} \\\\
		\textrm{with} \quad &H(u)^{T} \coloneqq [ \mathbb{I}_{p} \hspace{2mm} \nabla h(u)^{T}]\\
        \textrm{and}  \quad & w \coloneqq \begin{bmatrix}
			                 \Delta p\\ \Delta q
		                      \end{bmatrix}
\end{aligned}
\end{equation}
In this approach the gradient of the cost function $\Phi$ for the current iteration of OFO is projected onto the convex set of feasible control actions. The next vector of set points $\mathbf{u}(k+1) = [p_{1},...p_{j}, q_{1},...,q_{j}]^{T}$ is then calculated with:
\begin{equation}
    {\mathbf{u}(k+1) = \mathbf{u}(k) + \alpha\hat{\sigma}(u,y)}
\end{equation}
We are scaling the step-size by multiplying with $\alpha$ to tune and bound the system output and therefore ensure stability for convergence to the optimal solution of \eqref{eq:dispatch problem}. After actuation and a defined cycle time the controller continues with acquiring the grid state and updating the gradient of the cost function $\Phi$. As flexibility provision from sub-transmission level in this work is fulfilled in reaction to an external signal, the interaction between individual system operators is described in the next subsection.

\subsection{Hierarchical Control Stack}
\label{subsec:Hierarchy}
In order to participate in ancillary services, the flexibility from DER at sub-transmission level needs to be aggregated at the TSO-DSO-interface, i.e. the coupling transformer between EHV and HV layer. This can be generalized to the interfaces between flexibility providing and receiving systems, such as further underlying grid layers or Virtual Power Plants (VPPs). Additionally, it is worth noting that the decomposition between the grid layers can also be applied in cases where the transmission and sub-transmission systems are managed by a single system operator to decrease computational complexity in online operation. In the approach presented in this paper, this leads to the following interactions between the supervisory control at TSO-level and the flexibility control at DSO-level. First, the flexibility potential must be estimated by the DSO and made known to the TSO. Second, the TSO needs to request flexibility by sending set points for power flow to the DSO, denoted as $\mathbf{s}_{set}$. To track the requested set points the controller on sub-transmission level needs to update its cost function $\Phi$ upon receiving a new set point (see \eqref{eq:dispatch problem} and \eqref{eq:ofo_qp}). This results in the following sequence of steps for each iteration of the control stack:
\begin{enumerate}
    \item Acquire the current grid state by measurement of power flows and bus voltages.
    \item Update the gradient of the cost function $\nabla \Phi$ for the current step.
    \item Determine new set points for controllable flexibility by solving \eqref{eq:ofo_qp}.
    \item Update cost function $\Phi$ based on current set point for load flow at PCC $\mathbf{s}_{set}$.
\end{enumerate}

\section{Robust Trajectory Sets}
\label{sec:ROE}
To verify the controllers behavior we investigate its convergence and performance on the set of all technically feasible operating points. To this end, we define conditions that can be applied to the set of resulting trajectories from control actions, so called trajectory sets. For a discrete system trajectory described by a sequence $\{y(k)\}_{k=0}^{T}$ we define a trajectory set as follows.

\begin{definition}(Trajectory set)
A trajectory set for an initial state $y_{0}$ and a series of control inputs $u(k) \in \mathcal{U}$ as the set of all possible resulting trajectories $y(k)$ of a system with: 
\begin{equation}
\label{eq:def_oe}
\mathcal{E}(k) = \left\{ \mathbf{y}(k) \mid \mathbf{y}(k) = f(\mathbf{y}_0, \mathbf{u}(k), k), \ \mathbf{u}(k) \in \mathcal{U} \right\}
\end{equation}
\end{definition}

To validate the proposed control stack, we are analyzing the trajectory set for the final state $y(k_f)$. We define the final state of a trajectory based on the performance goals described in the cost function of the controller (see \eqref{eq:dispatch problem}). For the proposed control stack this can be described by reaching the requested set point for the complex interconnecting load flow $s_{\text{PCC}}$.

\subsection{Feasibility of System States}
\label{subsec:Feasibility}
In this work, we define the term \emph{Safety} in a control theory sense, i.e. as a system-set property of the power grid, considering its operational constraints, as a dynamical system. A formal definition is provided in \cite{Wang_2023}. We distinguish this from the term \emph{Security}, which is more commonly used in context of power systems and refers to the system’s ability to withstand perturbations \cite{Kirschen_2002}. To evaluate the safety of system states that result from actuation of a central controller we require a known safe set of such states considering the constraints of the system as a baseline to evaluate against.
\begin{definition}(Feasible operating region)
The feasible operating region (FOR) of a system is defined as the set of all load flows at the PCC $s_{\text{PCC}}$ that satisfy both equality and inequality constraints of the system:
\begin{equation}
\label{eq:def_for}
\mathcal{F} = \left\{ s_{\text{PCC}} \mid g(s_{\text{PCC}}, \mathbf{x}) = 0, \ h(s_{\text{PCC}}, \mathbf{x}) \leq 0 \right\}
\end{equation}
\end{definition}
To determine the FOR an approach based on sampling the hull of $\mathcal{F}$ is chosen. For a fixed angle $\vartheta \in [0^{\circ}, 360^{\circ})$ and with $\alpha,\beta \in \{-1,1\}$ we formulate the optimization problem for determining the FOR with:
\begin{equation}
\label{eq:for_det}
\begin{aligned}
    \min_{\mathbf{x}} &\quad \Phi =  -\alpha p_{\text{PCC}} -\beta q_{\text{PCC}} \\
    \text{s.t.} &\quad g(\mathbf{x}) = 0, \\
                      &\quad h(\mathbf{x}) \leq 0 \\
                      &\quad \tan{\vartheta} = \frac{q_{\text{PCC}}}{p_{\text{PCC}}} = \frac{\alpha}{\beta}
\end{aligned}
\end{equation}
In this approach we maximize the load flow for a fixed direction in the solution space, subject to equality constraints $g(\mathbf{x})$ and inequality constraints $h(\mathbf{x})$ of the AC-OPF problem. These include the full grid model with power flow equations in $g(\mathbf{x})$, as well as limits on bus voltages, branch flows, generators, and loads in $h(\mathbf{x})$. The FOR can subsequently be interpreted as a projection of the high-dimensional solution space of \eqref{eq:for_det} onto the two-dimensional PQ-plane at the PCC. This yields the theoretical operating range of the system considering all stationary limits of the detailed grid model. We are therefore able to use the FOR $\mathcal{F}$ to validate individual system states $\mathbf{y}(k)$ that result from the actuation of the controller with respect to these limits. As we choose the inequality constraints $g(\mathbf{x})$ in \eqref{eq:dispatch problem} to be identical to \eqref{eq:for_det} we can make the following assumption for system states resulting from control actions during flexibility provision. We denote the projection of the full system state $\mathbf{y}$ onto the FOR as $\mathbf{y}^*$.
\begin{assumption}(Feasibility of system states)
	\label{the:assumption_feas}
Any system state $\mathbf{y}(k)$ that results from actuation by the OFO controller is feasible subject to $g(\mathbf{x})$ and $h(\mathbf{x})$ if $\mathbf{y}^*(k) \in \mathcal{F}$.
\end{assumption}
Furthermore, we calculate the sensitivities used in the proposed OFO controller based on the same grid model we use to compute $\mathcal{F}$. A full grid model is therefore needed beforehand to perform this evaluation. Here we can distinguish between the operational planning phase and the actual online operation of the controller. In the operational planning phase, we use the full grid model to compute the FOR. During online operation, the controller relies on precomputed sensitivities, reducing the computational complexity. Using Assumption \autoref{the:assumption_feas}, we are able to validate individual system states, system trajectories and consequently trajectory sets that result from the actions of the proposed controller. In the next section we define conditions for the safeness of trajectory sets with respect to the system constraints using the FOR as reference for safe operation.
\begin{figure}[tb]
	\centering
	\includegraphics[width=1\linewidth]{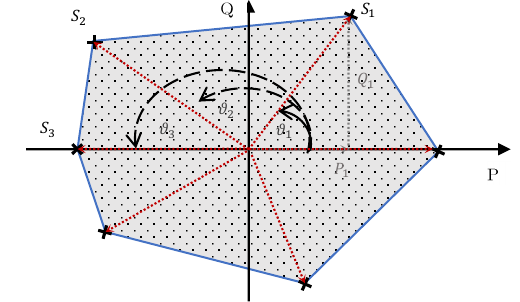}
	\caption{Angle-based determination of the FOR based on the AC optimal power flow.}
	\label{fig:angle_for}
\end{figure}
\subsection{Feasibility of System Trajectories}
\begin{figure*}[tb]
	\centering
	\includegraphics[width=\textwidth]{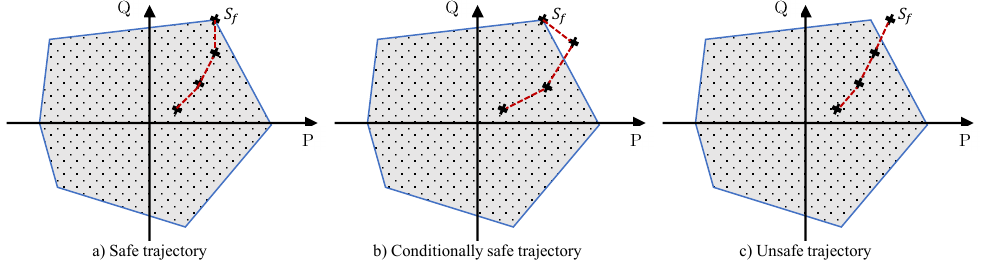} 
	\caption{Classification of possible trajectories resulting form actuation of system by OFO controller.}
	\label{fig:conditions}
\end{figure*}

We first define safeness for the trajectory set of a controller based on the constraint satisfaction during actuation of the system. We utilize the FOR as defined in \eqref{eq:def_for}. See \autoref{fig:conditions} for a graphical representation of the defined conditions.

\begin{definition}(Safe trajectory set)
A trajectory set \(\mathcal{E}\) is considered \emph{safe} if all states of all trajectories in $\mathcal{E}$ lie within the FOR \(\mathcal{F}\), i.e. the trajectory set is a subset of the FOR.
\begin{equation}
\label{eq:def_safe}
    \mathcal{E} \subseteq \mathcal{F}
\end{equation}
\end{definition}
Many constraints in power system operation can be interpreted as soft constraints as they do not immediately lead to critical situations, e.g. moderate over-loading of transmission lines does neither lead to immediate damage nor does it immediately trigger protection devices. As a temporary violation of constraints might be allowable, we give a supplementary condition for safeness that is based on evaluating constraint satisfaction for only the final converged steady state of the system. With the assumption that intermediate system states violate constraints within temporarily acceptable bounds we define a conditionally safe trajectory set as follows.
\begin{definition}(Conditionally safe trajectory set)
A trajectory set \(\mathcal{E}\) is considered \emph{conditionally safe} if all final states of all trajectories in \(\mathcal{E}\) lie within the FOR \(\mathcal{F}\).
\begin{equation}
\label{eq:def_consafe}
\{\mathbf{y}^*(k_f) \mid \mathbf{y}(k) = f(\mathbf{y}_0, \mathbf{u}(k), k), \ \mathbf{u}(k) \in \mathcal{U}\} \subseteq \mathcal{F}
\end{equation}
\end{definition}
For the proposed formulations, \emph{safeness} implies \emph{conditional safeness}. Lastly we define \emph{unsafe} trajectory sets based on constraint satisfaction of the final states of all trajectories $\mathbf{y}(k) \in \mathcal{E}$.
\begin{definition}(Unsafe trajectory set)
A trajectory set \(\mathcal{E}\) is \emph{unsafe} if there exists at least one trajectory \(\mathbf{y}(k)\) within \(\mathcal{E}\) such that its final state \(\mathbf{y}(k_f)\) lies outside the feasible operating region \(\mathcal{F}\).
\begin{equation}
\label{eq:def_unsafe}
\mathcal{E} \text{ is unsafe if } \exists \ \mathbf{y}(k) \in \mathcal{E} \ \text{:} \ \mathbf{y}^*(k_f) \notin \mathcal{F}
\end{equation}
\end{definition}
Using these conditions we are able to evaluate a controller that is subject to disturbances $d$ based on its resulting trajectory sets.
\subsection{Trajectory Sets under Uncertainty}
To specifically analyse the robustness of the control stack, we evaluate the performance based on the defined conditions for present uncertainties and noise. We first give a definition of robustness that utilizes the condition defined in \eqref{eq:def_safe}.
\begin{definition}(Robust control stack)
A control stack is robust for a set of unknown uncertainties and noise with known distributions if all resulting trajectory sets are a subset of the feasible operating region of the controlled grid.
\begin{equation}
\label{eq:def_robust}
\forall \mathbf{y}_0, \forall \mathbf{u}(k): \forall \mathcal{E}(\mathbf{y}_0, \mathbf{u}) \subseteq \mathcal{F},
\end{equation}
\end{definition}
Using this definition we evaluate a controller stack for a fixed formulation of the optimization problem, a fixed gain $\alpha$ for a set of uncertainties and noise. We describe the transition of system states from $\mathbf{y}_{k-1}$ to $\mathbf{y}_k$ by the control actions $\mathbf{u}(k-1)$ and by modeling the individual uncertainties where:
\begin{equation}
    \mathbf{y}(k) = h^*(\mathbf{u}) + \omega_l + \omega_y
\end{equation}
\paragraph{Uncertainty of Load}
In cases where we only assume controllability of generation units we consequently consider the current operating points of loads within the system as one of the main disturbances affecting the distribution grid during control of flexibility. We model the variation of the load over time according to \cite{Picallo_2020} for a load covariance matrix $\Sigma_l$ with: $\omega_{l} \sim \mathcal{N}(0, \Sigma_l)$. We differentiate the loads by type (households, industry and commercial) and adjust the entries of the covariance matrix accordingly. 

\paragraph{Measurement Noise}
As the proposed control architecture is using direct measurements of bus voltages and branch flows as feedback in the loop it is subject to any noise present in the respective signals. We model the noise term as uniform with $\omega_{y} \sim \mathcal{U}([a, b])$ uncorrelated for all measurements in $y$, in order to specifically evaluate the robustness of the proposed controller with respect to these bounds.

\paragraph{Mismatch of sensitivity map}
Additionally to the non-modeled dynamics of the system, i.e. uncertainty of load and noisy measurements, we consider a mismatch in model and physical system by introducing bounded errors to the steady-state sensitivity map $\nabla h(u)$. These are described by a bounded uniform distribution $\omega_s \sim \mathcal{U}([c,d])$. Similarly to the modeling of the measurement noise, we chose this distribution to provide a bounded approximation of the model mismatch and to be able to evaluate the proposed controller with respect to it. This results in the adjusted sensitivity map $\nabla h^*(u)$. Using the uncertainties and their respective distributions, we perform Monte Carlo simulations of the controller converging to reference values on the FOR. This allows for an evaluation of the performance and robustness of the controller in presence of disturbances. We are using the conditions we defined in previous subsections to validate the resulting trajectory sets. Furthermore, we evaluate the iterations needed for convergence to the requested load flows. The results of these case studies are detailed in the next section.

\section{Case Studies}
\label{sec:Case Studies}
In the following section we demonstrate the presented methods and conditions by applying them to an exemplary sub-transmission system. We detail the results of the individual case studies, that have been carried out for this work. First, we introduce the scenario used in all case studies. Second, the performance of the proposed controller is analyzed without disturbance. Third, the robustness and performance of the controller is evaluated subject to disturbances using the defined conditions.

\subsection{Scenario Description}
\label{subsec:Scenario}
\begin{figure}[tb]
	\centering
	\includegraphics[width=1\linewidth]{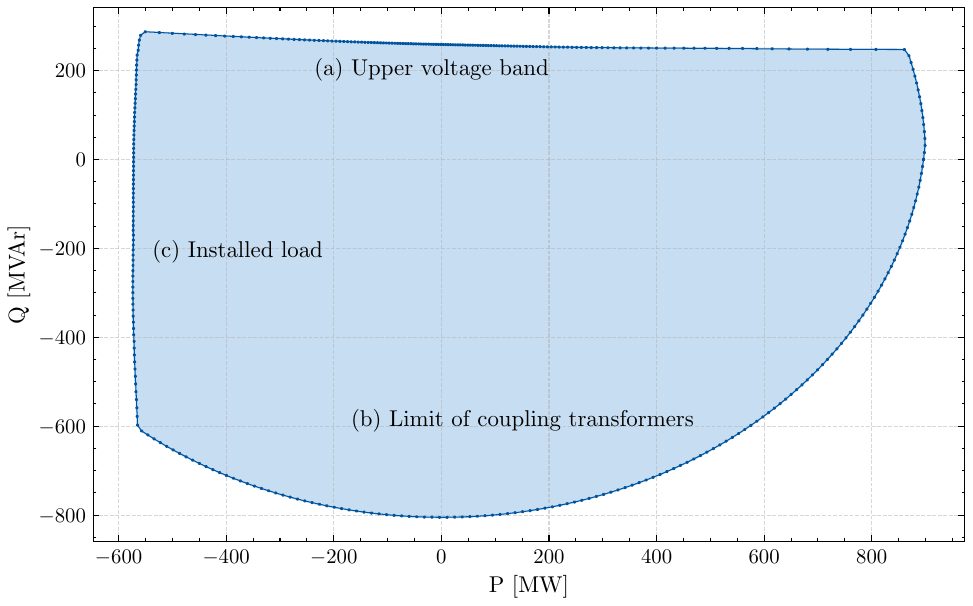}
	\caption{Feasible operating region of sub-transmission system, constrained by (a) the upper voltage band (upper limit), (b) the coupling transformer (lower and right-hand side limits) and (c) the installed load (left-hand side limit)}
	\label{fig:for_base}
\end{figure}

To showcase the presented control architecture, we analyze its behavior for an exemplary high-voltage sub-transmission system \cite{Meinecke_2020} by evaluating the conditions defined in the previous sections. Its FOR $\mathcal{F}$ is shown in \autoref{fig:for_base} for the PCC connecting it to the superimposed transmission system. The operational range of the sub-transmission grid is limited by different constraints for different regions of $\mathcal{F}$. As the controller behavior is directly dependent on the system constraints, the context of the bounds for the relevant system states is important to give context to all following results. The extreme points of the FOR are determined by the installed capacity of load and generation (see \autoref{tab:grid-parameters}) as well as limiting grid constraints.
\begin{table}[b]
	\centering
	\caption{Case Study: Grid Parameters \cite{Meinecke_2020}}
	\label{tab:grid-parameters}
	\begin{tabular}{@{}llllll@{}}
		\toprule
		\textbf{$V_{\text{N}}$} & \textbf{Buses} & \textbf{Lines} & \textbf{Load} & \textbf{DER} & \textbf{Voltage Band} \\ \midrule
		110 kV & 119 & 151 & 535.3 MVA & 1432.25 MVA & $V_{\text{N}}\pm 10\%$  \\
		\bottomrule
	\end{tabular}
\end{table}

\subsection{Trajectory Sets at Point of Common Coupling}
\label{subsec:case_1}
\begin{figure}[tb]
	\centering
	\includegraphics[width=1\linewidth]{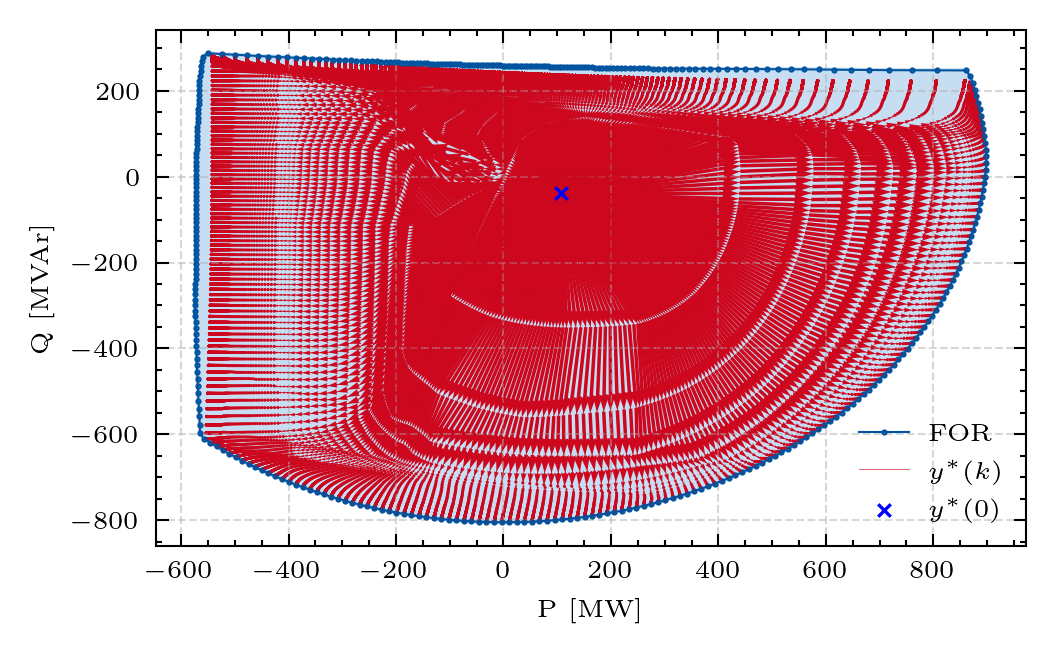}
	\caption{Convergence of controller to set points on hull of FOR.}
	\label{fig:traj}
\end{figure}

\begin{figure}[tb]
	\centering
	\includegraphics[width=1\linewidth]{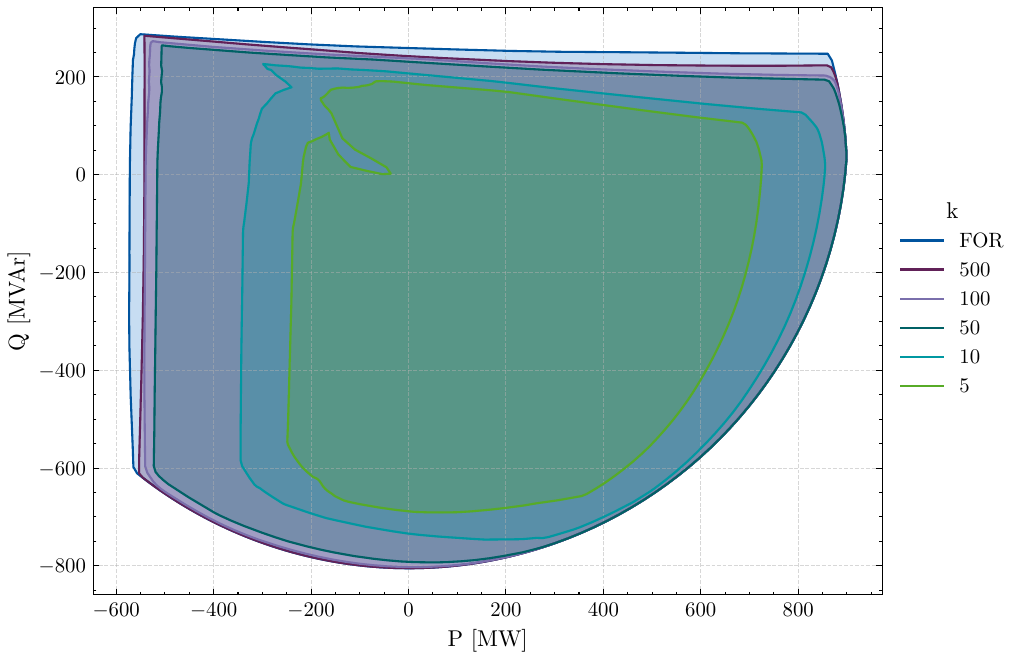}
	\caption{Development of trajectory sets over iterations on FOR without disturbance.}
	\label{fig:base_ts}
\end{figure}

To evaluate the controller's performance in the absence of disturbances, we analyze the resulting trajectories on the FOR for convergence to the vertices of the hull of $\mathcal{F}$. This analysis aims to evaluate the controller's convergence behavior across different regions of $\mathcal{F}$ and identify any constraints that significantly influence the results. \autoref{fig:traj} illustrates all resulting trajectories over 500 iterations on $\mathcal{F}$.
The convergence rate visibly varies across different regions of $\mathcal{F}$, as further demonstrated by the trajectory sets shown in \autoref{fig:base_ts}.
For the given initial operating point, 52\% of $\mathcal{F}$ is reachable within the first five iterations, primarily because no inequality constraints are active during these initial steps. The controller can therefore take larger steps along the current gradient of the cost function. 91\% of $\mathcal{F}$ is reached after 50 iterations with the convergence slowing down significantly afterwards. This also depends on the limiting constraints for the specific region of $\mathcal{F}$. Notably, in regions of $\mathcal{F}$ near the constraints of the coupling transformers, the set points are fully reachable within 10 to 50 iterations. However, on the left and top sides of $\mathcal{F}$, convergence takes significantly more iterations. After 500 iterations, the controller reaches 96\% of the FOR. The slower convergence in certain regions is attributed to the number of active constraints, such as voltage bands and load limits, that the OFO must account for. Consequently, two outer vertices of the FOR (bottom-left and top-right) remain unreachable within 500 iterations under the given controller tuning. The slower convergence observed for the left and top regions of $\mathcal{F}$ is likely due to the high number of individual active constraints in these areas, which restrict the controller's ability to take larger steps. Overall, all resulting trajectory sets $\mathcal{E}(k)$ remain \emph{safe} in accordance with \eqref{eq:def_consafe} for all iterations $k$. The controller and its tuning are therefore robust in the sense of \eqref{eq:def_robust}. The trajectory set for the first five iterations $\mathcal{E}(5)$, shown in \autoref{fig:base_ts}, demonstrates slower convergence around a load flow of $s_{\text{PCC}} = 0 \ \text{MVA}$. This is likely due to the fact that the controller approaches the lower bound of zero of loading of the coupling transformers for the given scenario. Therefore if the inequality constraints are evaluated against absolute or relative values the performance may temporarily decline even without approaching critical system states.

\subsection{Influence of Disturbance}
\begin{figure}[tb]
	\centering
	\includegraphics[width=1\linewidth]{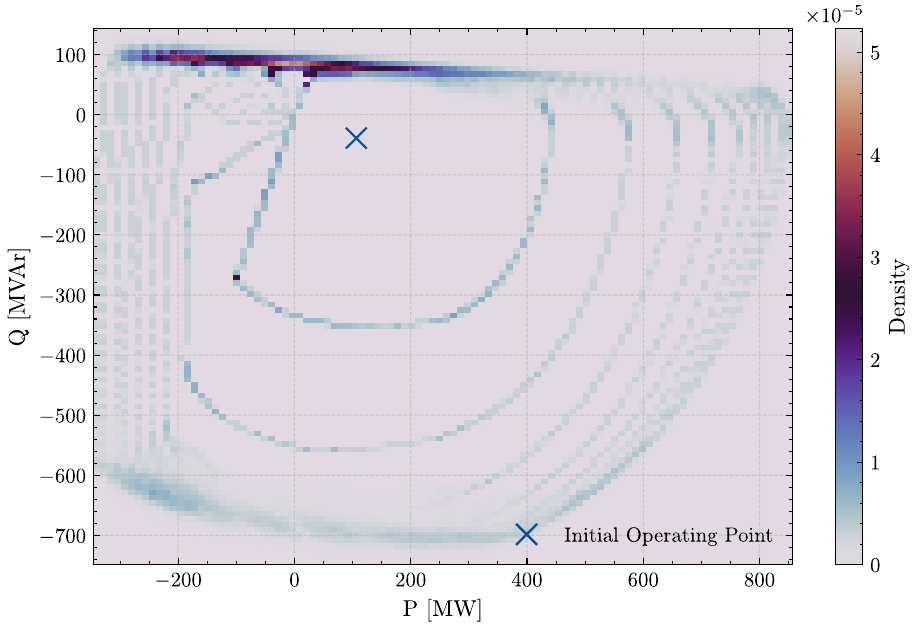}
	\caption{Density of system states on the FOR for the first ten control actions.}
	\label{fig:density_steps}
\end{figure}

\begin{figure}[tb]
	\centering
	\includegraphics[width=1\linewidth]{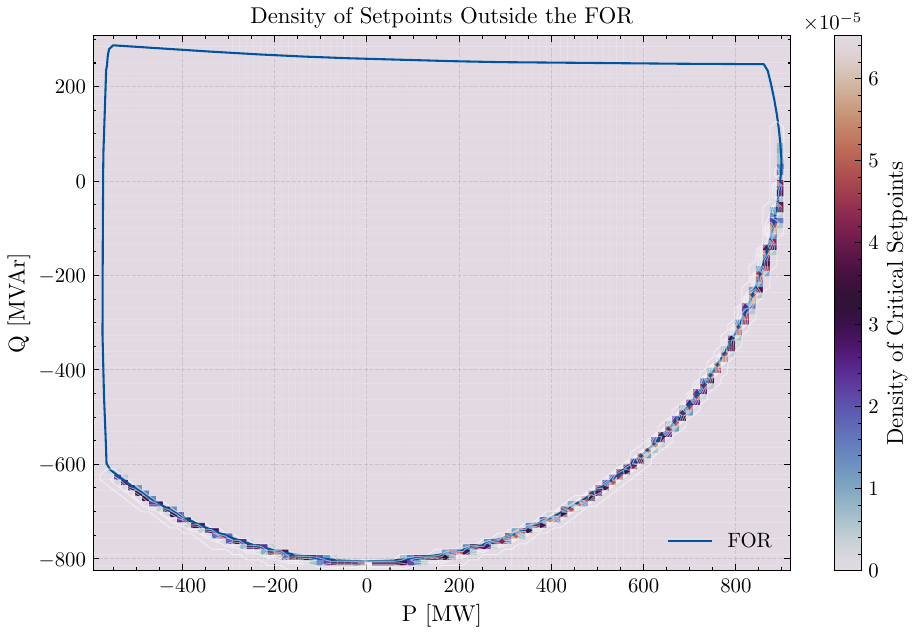}
	\caption{Density of critical system states on the FOR.}
	\label{fig:density_crit}
\end{figure}

\begin{figure}[t]
	\centering
	\includegraphics[width=1\linewidth]{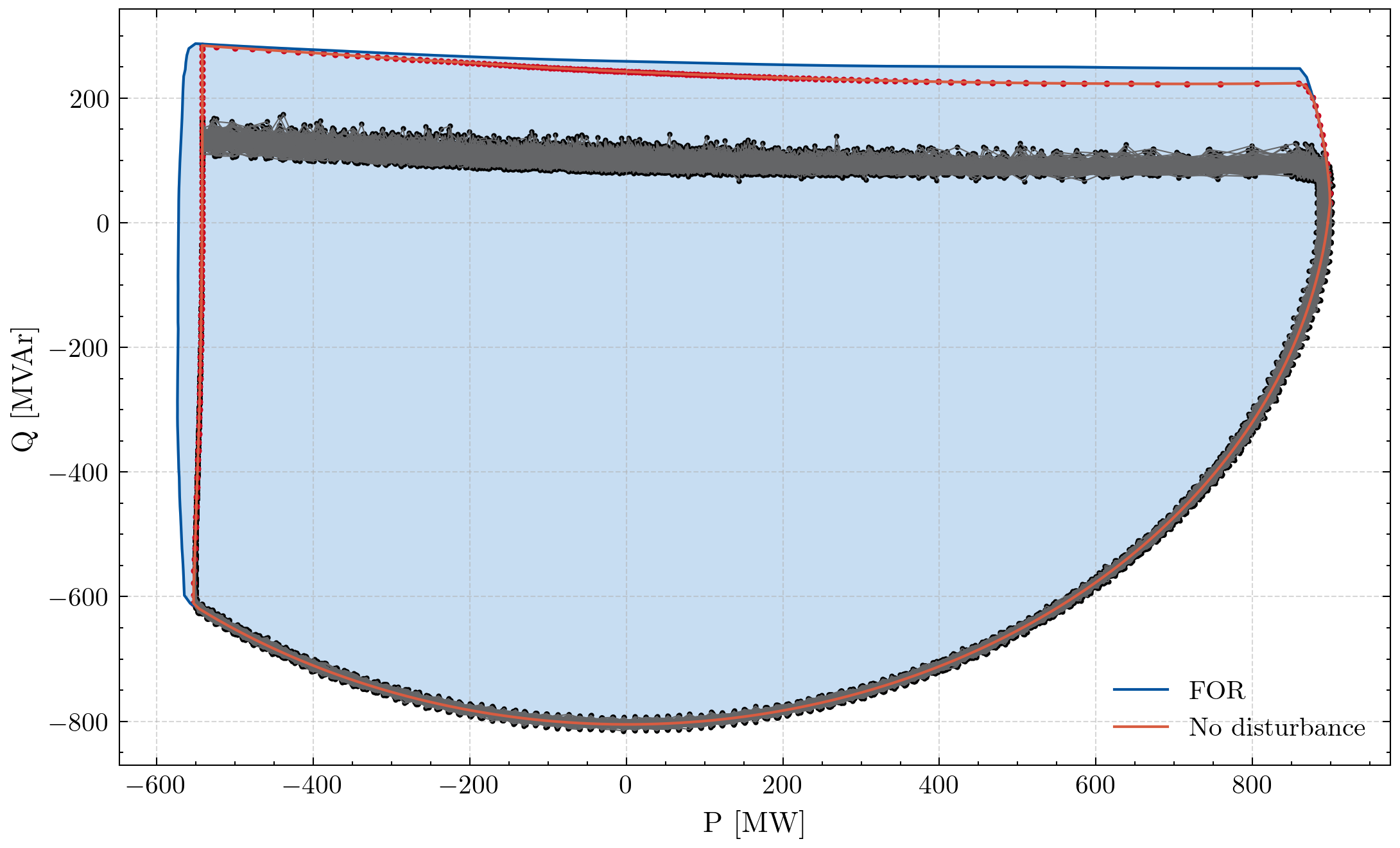}
	\caption{Final trajectory sets under on FOR subject to disturbances.}
	\label{fig:uncertain_traj}
\end{figure}

\begin{figure}[tb]
	\centering
	\includegraphics[width=1\linewidth]{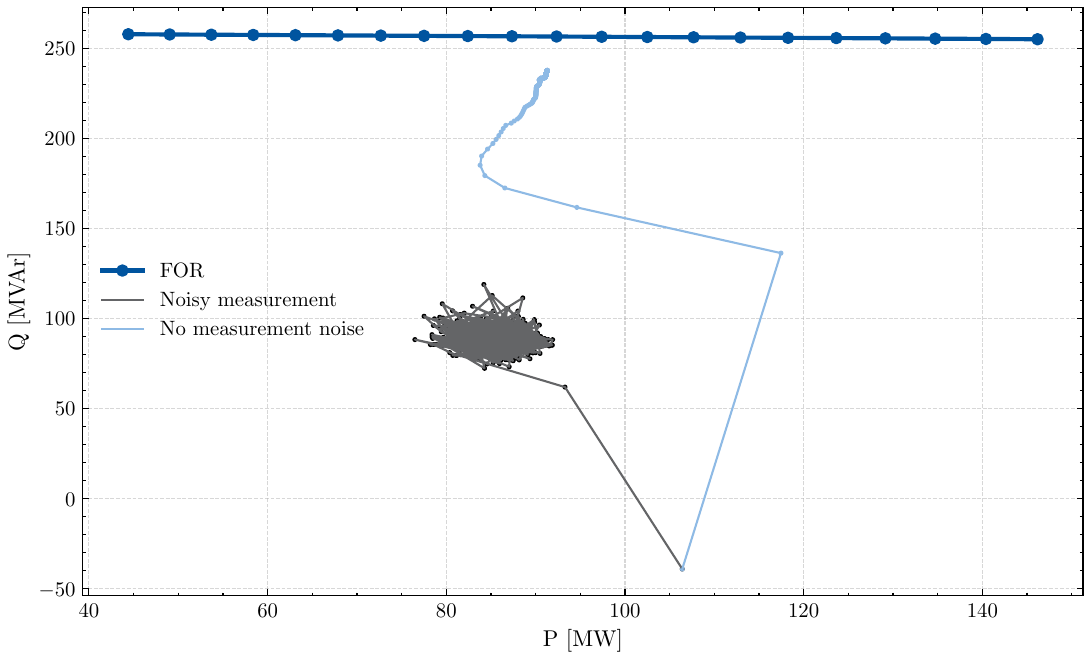}
	\caption{Impact of noisy measurements on controller performance.}
	\label{fig:single_uncertain_traj}
\end{figure}

\begin{figure}[t]
	\centering \includegraphics[width=1\linewidth]{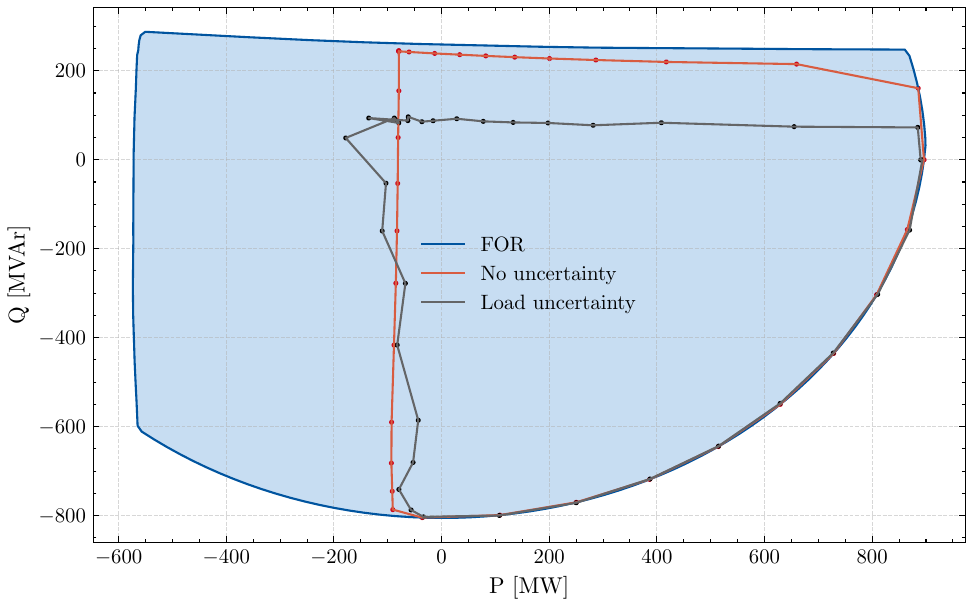} \caption{Exemplary final trajectory set considering uncertainty in load.} \label{fig:traj_set_no_load}
\end{figure}

This case study examines the behavior of trajectory sets under uncertainty and noisy measurements, focusing on the high-voltage sub-transmission grid described in a previous subsection. The analysis evaluates trajectories that result from controlling load flow at the PCC to the transmission system. Parameter uncertainty, $\omega_s$, is modeled as a uniform distribution with bounds of $\pm 5\%$, while measurement noise, $\omega_y$, is similarly modeled with bounds of $\pm 2\%$ on all entries in $y$. First, we consider both load and generation of the system to be fully controllable. The controller is evaluated by performing in total 63.000 simulations of the trajectories with 500 iterations of OFO each under varying disturbances. Based on the resulting system states we can evaluate the density of the different load flows resulting from the control actions at the coupling transformers on $\mathcal{F}$ as shown in \autoref{fig:density_steps}. We define the density for individual bins on $\mathcal{F}$ based on the number of resulting system states in a bin $n_{ij}$ and the area of the bin $a_{ij}$ with:
\begin{equation}
	\rho(p, q) = \frac{n_{ij}}{a_{ij} \cdot n_{\text{total}}}
\end{equation}
The density clearly reflects the general development of the trajectory sets for the controller as shown in \autoref{fig:base_ts}, indicating that even in when subjected to uncertainties the control actions follow a clear trajectory. It is observable that the density is decreasing for later iterations of the controller. This is explainable by the increasing number of constraints that can become active as the system operates closer to its bounds.
\autoref{fig:density_crit} illustrates the density of system states lying outside the FOR.
Approximately 40\% of all trajectories within the resulting sets contain critical system states. By projecting the overall system states onto the FOR, we identify the limiting constraints under uncertainty. As shown in \autoref{fig:density_crit}, critical system states are primarily distributed along the lower right section of the FOR, beginning at $\underline{s}_{PCC} = -500 \ \text{MW} -j600 \ \text{MVAr}$. This distribution is attributed to the controller's behavior in response to active limiting constraints during operation. Specifically, the constraints imposed by the PCC transformers, as described in the scenario, limit the load flow in this region of the FOR. The observed distribution of critical states suggests that the controller struggles to accommodate the combined effects of parameter uncertainty and transformer constraints in the lower right section of the FOR. A detailed analysis of the sensitivity of the system states to $\omega_s$ and $\omega_y$ confirms that deviations in measurements significantly affect the load flow trajectory, especially near the constraint-dominated regions. As shown in \autoref{fig:uncertain_traj} the final trajectory sets under uncertainty vary significantly from the final set without disturbances.
Especially, the achievable trajectory set of load flows for reactive power provision (top bound of $\mathcal{F}$) to the superimposed system is notably smaller than in the previous case study. This again can be attributed to the density of active constraints in that region of $\mathcal{F}$. As the individual upper bus voltage limits are nearly reached, the noisy measurement leads OFO to perceive these constraints as active. As the noise of the individual sensors is not correlated this leads to a non-convergence of trajectories as OFO is reacting simultaneously to varying perceived constraint violations within the range of the measurement noise $\omega_{y}$. This is further illustrated in \autoref{fig:single_uncertain_traj} showing an exemplary trajectory with and without noisy measurements.
This highlights the difference in influence of measurement uncertainty and sensitivity mismatch as the controller without measurement noise is able to converge even with an augmented sensitivity map. This aligns with previous theoretical considerations guaranteeing convergence of projected gradient descent to an optimal solution for a sufficiently small gain $\alpha$ \cite{Hauswirth_2024}. To further validate the proposed methods, we analyze a scenario without load controllability and present the corresponding final trajectory set in \autoref{fig:traj_set_no_load}. In this case, the controller attempts to converge to the vertices of the Feasible Operating Region (FOR) with a reduced resolution of $\mathcal{F}$. The reduced availability of controllable power is clearly reflected in the size of the trajectory set, which is significantly smaller compared to the scenario with full controllability of all actuators. The uncertain behavior of the loads introduces additional variability, as depicted in the final trajectory set (gray), diverging from the base scenario without uncertainty. Despite this, the control stack demonstrates robustness as defined in \eqref{eq:def_robust}, with no observed violations of system limits across all evaluated states. This result highlights the adaptability of the proposed control framework under constrained controllability and uncertain system conditions. While the reduced flexibility imposes stricter limitations on achievable trajectory sets, the controller successfully maintains safe operation by dynamically adjusting to the available flexibility and uncertainties.

\section{Conclusion}
\label{sec:Conclusion}
Online feedback optimization (OFO) is discussed as a viable approach for control problems in power system operation. We proposed a cascaded control architecture based on OFO to facilitate the participation of sub-transmission-level flexibility in ancillary services such as congestion management. The cascaded architecture facilitates efficient coordination between system operators as it decomposes the dispatch problems along the system boundaries. Robustness and performance of the controller were systematically evaluated through set-based considerations of the system's Feasible Operating Region (FOR). Case studies demonstrated that the controller effectively tracks load flow set points at the coupling transformer within a significant portion of the FOR in undisturbed scenarios. Convergence was slower near system bounds, where constraints become more active. Under uncertainty, controller performance was shown to degrade, particularly in the presence of noisy measurements, which introduced variability in constraint activation. Despite these challenges, the OFO controller maintained safety and robustness consistent with theoretical guarantees. The proposed OFO-based framework offers significant advantages for the online operation of power systems. It is particularly effective in scenarios and use-cases, such as curative system operation or online voltage support, where exact system models are unavailable and where computational efficiency is crucial to avoid critical system states. 

\section*{Acknowledgment}
\begin{wrapfigure}{r}{0.13\textwidth}
  \vspace{-\baselineskip}
  \vspace{-\baselineskip}
  \includegraphics[width=0.13\textwidth]{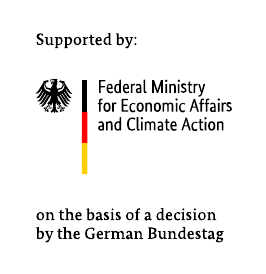}
  \vspace{-\baselineskip}
    \vspace{-\baselineskip}
        \vspace{-\baselineskip}
\end{wrapfigure}

This project received funding from the German Federal Ministry for Economic Affairs and Climate Action under the agreement no. 03EI4046E (PROGRESS).

Computations were performed with computing resources granted by RWTH Aachen University under project rwth1703.

\vspace{12pt}
\color{red}

\end{document}